\documentclass[aps,prb,twocolumn,superscriptaddress,showpacs]{revtex4}
\usepackage{graphicx} 

\begin{document}

\title{Single crystal $^{31}$P NMR studies of the frustrated
square lattice compound Pb$_{2}$(VO)(PO$_{4}$)$_{2}$}

\author{R. Nath}
\altaffiliation[Present address: ]{School of Physics, Indian Institute 
of Science Education and Research, Thiruvananthapuram-695016, India.}
\affiliation{Ames Laboratory and Department of Physics and Astronomy, 
Iowa State University, Ames, IA 50011, USA}
\affiliation{Max Planck Institut f\"{u}r Chemische Physik fester
Stoffe, N\"{o}thnitzer Str.~40, 01187 Dresden, Germany}
\author{Y. Furukawa}
\affiliation{Ames Laboratory and
Department of Physics and Astronomy, Iowa State University, Ames, IA
50011, USA}
\author{F. Borsa}
\affiliation{Ames Laboratory and
Department of Physics and Astronomy, Iowa State University, Ames, IA
50011, USA}
\affiliation{Dipartimento di Fisica ``A. Volta," Universit\`{a} di 
Pavia, I-27100 Pavia, Italy}
\author{E. E. Kaul}
\author{M. Baenitz}
\author{C. Geibel}
\affiliation{Max Planck Institut f\"{u}r Chemische Physik fester
Stoffe, N\"{o}thnitzer Str.~40, 01187 Dresden, Germany}
\author{D. C. Johnston}
\affiliation{Ames Laboratory and
Department of Physics and Astronomy, Iowa State University, Ames, IA
50011, USA}

\date{\today }

\begin{abstract}
The static and dynamic properties of V$^{4+}$ spins ($S=1/2$) in the 
frustrated square lattice compound Pb$_{2}$(VO)(PO$_{4}$)$_{2}$ were 
investigated by means of magnetic susceptibility $\chi$ and 
$^{31}$P nuclear magnetic resonance (NMR) shift ($K$) and $^{31}$P nuclear 
spin-lattice relaxation rate $1/T_{1}$ measurements on a single crystal.
This compound exhibits long-range antiferromagnetic order below 
$T_{\rm N} \simeq 3.65$~K\@. NMR spectra above $T_{N}$ show two distinct 
lines corresponding to two inequivalent P-sites present in the crystal 
structure. The observed asymmetry in hyperfine
coupling constant for the in-plane (P1) P-site directly points towards 
a distortion in the square lattice at the microscopic level, consistent 
with the monoclinic crystal structure. The nearest- and next-nearest-neighbor 
exchange couplings were estimated by fitting $K$ versus temperature 
$T$ by a high temperature series expansion for the spin susceptibility 
of the frustrated square lattice to be $J_{1}/k_{\rm B}=(-5.4 \pm 0.5)$~K
(ferromagnetic) and $J_{2}/k_{\rm B} = (9.3 \pm 0.6)$~K (antiferromagnetic),
 respectively. $1/(T_{1}T\chi)$ is almost $T$-independent at high 
 temperatures due to random fluctuation of spin moments. Below 20~K, 
 the compound shows an enhancement of $1/(T_{1}T\chi)$ which arises 
 from a growth of antiferromagnetic spin correlations above $T_{\rm N}$.
 Below $T_{\rm N}$ and for the field applied along the $c$-axis, the NMR 
 spectrum for the P1 site splits into two satellites and the spacing 
 between them increases monotonically with decreasing $T$ which is a 
 direct evidence of a columnar antiferromagnetic ordering with spins 
 lying in the $ab$-plane.  This type of magnetic ordering is consistent with expectation from the $J_{2}/J_{1}\simeq -1.72$ ratio.  The critical exponent 
$\beta=0.25 \pm 0.02$ estimated from the temperature dependence 
of the sublattice magnetization as measured by $^{31}$P NMR at 11.13~MHz 
is close to the value (0.231) predicted for the two-dimensional XY model.
\end{abstract}


\maketitle

\section{\textbf{Introduction}}
\begin{figure}
\includegraphics [width=3.2in] {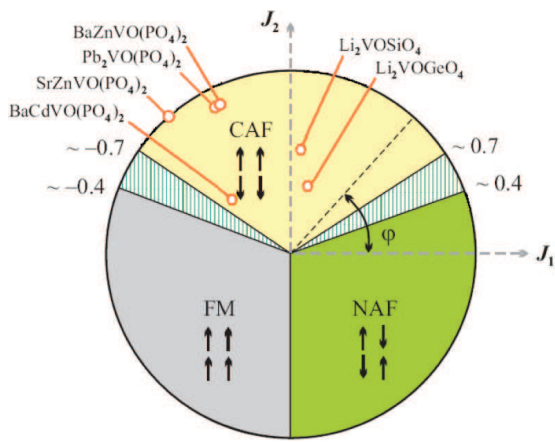}
\caption{\label{phasediagram} (Color online) The $J_{1}-J_{2}$ phase 
diagram of the two-dimensional spin-1/2 square spin lattice showing 
different ordered phases and the quantum spin liquid (QSL) regimes. 
The two QSL regimes are the two vertically hatched areas. The 
boundaries of the QSL regimes are not precisely known. The other 
regions of the phase diagram contain the ferromagnetic 
(FM), columnar antiferromagnetic (CAF), and N\'{e}el antiferromagnetic
(NAF) phases. The location of all the investigated compounds so far 
in the columnar antiferromagnetic (CAF) phase are shown.\cite{kaul2004,
kaul2005,nath2008,rosner2003} The radius of the circle is 12.2~K\@. 
Due to relatively large exchange couplings, PbVO$_{3}$ and VOMoO$_{4}$ are 
not included in the phase diagram, for which $J_{1}=203$~K and 
$J_{2}=77$~K (Ref.~\onlinecite{tsirlin2008}) and $J_{1}=110$~K and 
$J_{2}=22$~K,\cite{carretta2002b} respectively. (After 
Refs.~\onlinecite{nath2008} and \onlinecite{skoulatos2009})}
\end{figure}

Understanding the ground state properties of frustrated low-dimensional
spin systems is a central issue in current condensed matter physics. These
systems have competing interactions which stabilize different states
with distinct symmetries.\cite{misguich2004} An interesting phenomenon
in spin systems is the formation of a spin-liquid with no long-range magnetic
order (LRO) due to suppression by geometric/magnetic frustration. 
A simple example is the frustrated $S=1/2$ square
lattice (FSL) model. In this model which is also known as $J_{1}-J_{2}$
model, the spin Hamiltonian is 
\begin{equation}
{\cal H}=(k_{\rm B}J_{1}) \sum\limits_{<ij>} 
\vec{S_{i}} \cdot \vec{S_{j}}+(k_{\rm B}J_{2}) \sum\limits_{<ik>}
\vec{S_{i}} \cdot \vec{S_{k}},
\end{equation}
 where the first sum is over 
nearest-neighbor spin pairs and the second is over 
next-nearest-neighbor spin pairs. A positive $J$ corresponds 
to an antiferromagnetic (AF) exchange interaction and a negative 
$J$ to a ferromagnetic one. In this paper, $J_{1}$ and $J_{2}$ 
are expressed in temperature units, and $k_{\rm B}$ is 
Boltzmann's constant. The nearest-neighbor (NN) interaction $J_{1}$ 
along the side of the square can compete with the next-nearest-neighbor (NNN) interaction $J_{2}$ along the
diagonal of the square if, e.g., both are AF\@. Recently a 
generalized phase diagram including both antiferromagnetic 
and ferromagnetic $J_{1}$ and $J_{2}$ has been proposed
theoretically based on the frustration ratio $\alpha=J_{2}/J_{1}$ or
frustration angle $\phi={\rm tan}^{-1}(J_{2}/J_{1})$ as shown in 
Fig.~\ref{phasediagram}.\cite{shannon2004,shannon2006,schmidt2006}
It has three different ordered phases: ferromagnet [FM, wave vector
$\mathbf{Q}=(0,0)$], N\'{e}el antiferromagnet [NAF, $\mathbf{Q}=(\pi,\pi)$],
and columnar antiferromagnet [CAF, $\mathbf{Q}=(\pi ,0)$ or $(0,\pi)$].
The critical regions with disordered ground states are predicted to
occur at the boundaries NAF--CAF and the CAF--FM which are quantum 
spin liquid (QSL) phases. A gapless nematic state is suggested for 
$\alpha \sim -0.5$,\cite{shannon2006} while different
dimer phases (including resonating-valence-bond-type ones) are claimed to
exist for $\alpha $ close to 0.5.\cite{sushkov2001,sorella1998,capriotti2000,capriotti2001}
In addition to its intrinsic interest, a better understanding of 
the $J_{1}-J_{2}$ model is expected to play a vital role in 
understanding the magnetism in recently discovered FeAs-superconducting 
parent compounds.\cite{si2008,chen2008,kamihara2008}

Experimentally, however, few compounds satisfying the $J_{1}-J_{2}$ 
square spin lattice model have been realized so far.
Li$_{2}$VO$X$O$_{4}$ ($X$ = Si, Ge)\cite{melzi2001,melzi2000} 
are two well-studied compounds with antiferromagnetic $J_{1}$ 
and $J_{2}$ and lie in the CAF state of the phase diagram, far 
away from the QSL regime.\cite{rosner2002,rosner2003} Recently 
two new perovskite type compounds (Cu$X$)LaNb$_{2}$O$_{7}$ ($X$
= Cl, Br) were claimed to realize the $J_{1}-J_{2}$ model with
ferromagnetic $J_{1}$ and antiferromagnetic $J_{2}$.
\cite{kageyama2005,oba2006} However the validity of the 
frustrated square lattice model for these two systems is still
unclear.\cite{yoshida2007} Recent band structure calculations 
suggested that despite the layered structure 
of (CuCl)LaNb$_{2}$O$_{7}$, the spin system has rather pronounced
 one-dimensional character.\cite{tsirlin2009d} The compounds VOMoO$_{4}$
(Refs.~\onlinecite{carretta2002b} and \onlinecite{bombardi2005})
and PbVO$_{3}$ (Ref.~\onlinecite{tsirlin2008}) have been 
studied in light of the $J_{1}-J_{2}$ model
and were deduced to lie in the NAF region of the phase diagram.

The vanadium phosphates $AA^{'}$VO(PO$_{4}$)$_{2}$ ($AA^{'}$ 
= Pb$_{2}$, SrZn, BaZn, and BaCd) present another realization 
of the $J_{1}-J_{2}$ model with ferromagnetic $J_{1}$ and 
antiferromagnetic $J_{2}$ and fall in the CAF region of the
phase diagram close to the QSL regime.\cite{kaul2004,kaul2005,
nath2008,tsirlin2009a}
The magnetic properties of these compounds have been recently 
investigated by means of magnetization,\cite{kaul2004,tsirlin2009a} heat 
capacity,\cite{kaul2005,nath2008} neutron scattering,
\cite{skoulatos2007,skoulatos2009} and muon spin resonance
\cite{carretta2009} ($\mu$SR) experiments.
All these measurements were done on polycrystalline materials. 
However a clear picture about the spin dynamics in these compounds 
requires measurements on single crystals. Here we present
$^{31}$P NMR and magnetic susceptibility $\chi$ measurements on a 
large single crystal of Pb$_{2}$VO(PO$_{4}$)$_{2}$.
\begin{figure}
\includegraphics [width=3in]{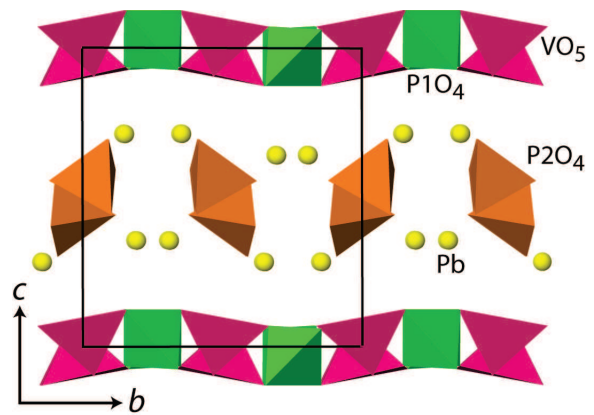}\\
\vspace{0.3in}
\includegraphics [width=2.5in]{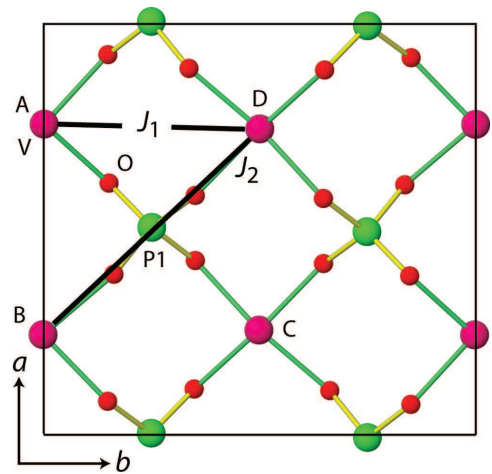}
\caption{\label{structure} (Color online) Upper panel: schematic diagram of
[VOPO$_{4}$] layers in Pb$_{2}$(VO)(PO$_{4}$)$_{2}$ running parallel 
to the $ab$-plane. Lower panel: projection onto the $ab$-plane of the 
unit cell showing the arrangement of VO$_{5}$ pyramids and P1O$_{4}$ 
tetrahedra in the plane. The approximate square lattice formed by 
four V$^{4+}$ spins (labeled as A, B, C, and D) is also shown indicating 
the nearest-neighbor ($J_{1}$) and next-nearest-neighbor ($J_{2}$) 
exchange interactions.}
\end{figure}

Pb$_{2}$(VO)(PO$_{4}$)$_{2}$ crystallizes in a monoclinic structure
with space group $P2_{1}/a$ and lattice parameters $a=8.747(4)$~\AA,
$b=9.016(5)$~\AA, $c=9.863(9)$~\AA, and $\beta=
100.96(4)^{\circ}$.\cite{shpanchenko2006} The structure
of Pb$_{2}$(VO)(PO$_{4}$)$_{2}$, presented in the upper panel 
of Fig.~\ref{structure}, is composed of [VOPO$_{4}$] layers 
extending parallel to the $ab$-plane which are separated along 
the $c$-axis by Pb atoms and isolated PO$_{4}$ tetrahedra as shown. 
The layers are slightly modulated due to lack of tetragonal symmetry. 
It is important to mention here that there are two inequivalent 
phosphorus sites (P1 and P2) present in the crystal structure. 
The [VOPO$_{4}$] layers are formed by corner-sharing V$^{4+}$O$_{5}$ 
pyramids with P1O$_{4}$ tetrahedra while the other site forms 
isolated P2O$_{4}$ tetrahedra lying between the [VOPO$_{4}$] 
layers. Thus the P1 site is expected to be strongly coupled 
and the P2 site weakly coupled to the V$^{4+}$ ($3d^{1}, S=1/2$) 
spins. In the lower panel of Fig.~\ref{structure} a 
section of the $ab$-plane is shown. In the approximate V$_{4}$ 
square A-B-C-D shown, the V$^{4+}$ spins interact
via a V-O-P1-O-V pathway involving two inequivalent oxygen atoms. 
The distances and angles of the pathways between the different V 
spins are also different which reflects the inequivalent superexchange 
paths. We have listed in Table~\ref{square} the distances between 
the pairs of four V$^{4+}$ spins A, B, C, D, and the angles between 
them. Small differences are seen in the distances along 
the edges, the angles between the edges, and the distances along 
the diagonals from square symmetry, indicating a small distortion of 
the square.

\begin{table}
\caption{\label{square} The distance between pairs of V$^{4+}$ spins 
along the edges and along the diagonals, and the interior angle between 
them, for the approximately square lattice shown in the lower panel of 
Fig.~\ref{structure} for Pb$_{2}$VO(PO$_{4}$)$_{2}$.}
\begin{ruledtabular}
\begin{tabular}{ccccccc}
  V-V pair & AB & BC & CD & DA & AC & BD \\ \hline
 Edges (\AA) & 4.669 & 4.665 & 4.669 & 4.410 &  &  \\ \hline
 Diagonals (\AA) &  &  &  &  & 6.267 & 6.295 \\ \hline
  & $\angle{\rm ABC}$ & $\angle{\rm BCD}$ & $\angle{\rm CDA}$ & $\angle{\rm DAB}$ &  &  \\ \hline
 Angles ($^{\circ}$) & 84.34 & 84.82 & 87.25 & 87.75 &  &  \\
\end{tabular}
\end{ruledtabular}
\end{table}

The exchange interactions $J_{1}$ and $J_{2}$ between V spins-1/2 
in Pb$_{2}$VO(PO$_{4}$)$_{2}$ were reported to be $-6.0$~K and 
$9.8$~K, respectively, and the compound was reported to undergo 
antiferromagnetic ordering at $T_{\rm N}\simeq 3.65$~K, likely 
of CAF type.\cite{kaul2004} Recently Tsirlin and
Rosner\cite{tsirlin2009} pointed out theoretically from band 
structure calculations that due to the reduced symmetry 
the $J_{1}$ and $J_{2}$ values within the V$_{4}$ square are
multivalued: ($J_{1}^{'}$, $J_{1}^{''}$) and ($J_{2}^{'}$, $J_{2}^{''}$),
 respectively and the experimental estimates of $J_{1}$ and 
 $J_{2}$ should be considered as averaged NN and NNN couplings: 
 $\bar{J_{1}}= (J_{1}^{'}+J_{1}^{''})/2$ and $\bar{J_{2}}=
 (J_{2}^{'}+J_{2}^{''})/2$.
From neutron scattering experiments, a diffuse magnetic background
was observed in the paramagnetic state at $T=20$~K which persists
well below $T_{\rm N}$. The ground state was suggested to be CAF 
type with a reduced ordered moment ($\mu \simeq 0.5\ \mu_{B}$/V$^{4+}$), 
where $\mu_{\rm B}$ is the Bohr magneton.\cite{skoulatos2007,
skoulatos2009} For a spin-1/2 with $g$-factor $g=2$, one would 
expect an ordered moment $\mu=gS \mu_{\rm B}=1\ \mu_{\rm B}$ per V atom. 
The diffuse background and the reduced ordered moment may be 
indications of quantum disorder predicted theoretically near 
the QSL regime.\cite{shannon2006} $\mu$SR experiments also
show a broad distribution of field at the muon site below 
$T_{\rm N}$ suggesting possible structural and/or magnetic 
disorder.\cite{carretta2009} The temperature dependence of the 
longitudinal muon relaxation rate above $T_{\rm N}$ and of the order 
parameter below $T_{\rm N}$ agree with expectations for the 2D XY model.
Both neutron scattering and $\mu$SR experiments were performed
on powder samples. In unpublished work, the $\chi(T)$ of a single 
crystal above $T_{\rm N}$ was found to be isotropic along different 
orientations.\cite{kaul2005} Below $T_{\rm N}$, it is
strongly anisotropic with the $b$-axis being the easy axis.\cite{kaul2005}

Nuclear magnetic resonance (NMR) is a very powerful experimental
tool to extract microscopic information at the individual sites of a
crystal. The technique has been applied to many transition metal
oxides and has played a significant role in elucidating their
microscopic magnetic characters. In this paper, we report a detailed
experimental investigation of static and dynamic properties of V$^{4+}$
spins in Pb$_{2}$(VO)(PO$_{4}$)$_{2}$ by means of $^{31}$P NMR and 
$\chi$ measurements. We have carried out $^{31}$P-NMR not only on 
a polycrystalline sample but also on a large single crystal in order 
to shed light on the spin structure and possible quantum disorder in 
Pb$_{2}$(VO)(PO$_{4}$)$_{2}$.

\section{\textbf{Experimental}}

Syntheses of the polycrystalline sample and of the crystal are reported
in Refs.~\onlinecite{carretta2009} and \onlinecite{shpanchenko2006},
respectively. The NMR measurements were carried out using pulsed NMR 
techniques on $^{31}$P (nuclear spin $I=1/2$ and gyromagnetic ratio 
$\gamma_{N}/2\pi = 17.237$ MHz/T) nuclei in the temperature range 1.5~K
$\leq T \leq $ 300~K. Laue x-ray back-scattering measurements previously 
determined in Ref.~\onlinecite{kaul2005} the orientation of the crystal 
axes with respect to the crystal faces for the crystal on which NMR 
measurements were performed. We have done the
NMR measurements at two different radio frequencies of $70$~MHz
and $11.13$~MHz which correspond to an applied field of about $4.06$~T and
$0.65$~T, respectively. Spectra were obtained either by Fourier
transform of the NMR echo signals or by sweeping the field. The
NMR shift $K=(\nu-\nu_{\rm ref})/\nu_{\rm ref}$ was determined by
measuring the resonance frequency of the sample ($\nu$) with respect
to nonmagnetic reference H$_{3}$PO$_{4}$ (resonance frequency
$\nu_{\rm ref}$). The $^{31}$P spin-lattice relaxation rate
$1/T_{1}$ was measured by the conventional single saturation pulse method.
For the analysis of NMR data, we measured the magnetic susceptibility $\chi(T)$
of the single crystal at 4~T for field applied along $c$-axis in a commercial 
(Quantum Design) SQUID (Superconducting Quantum Interference Device) magnetometer.

\begin{figure}
\includegraphics {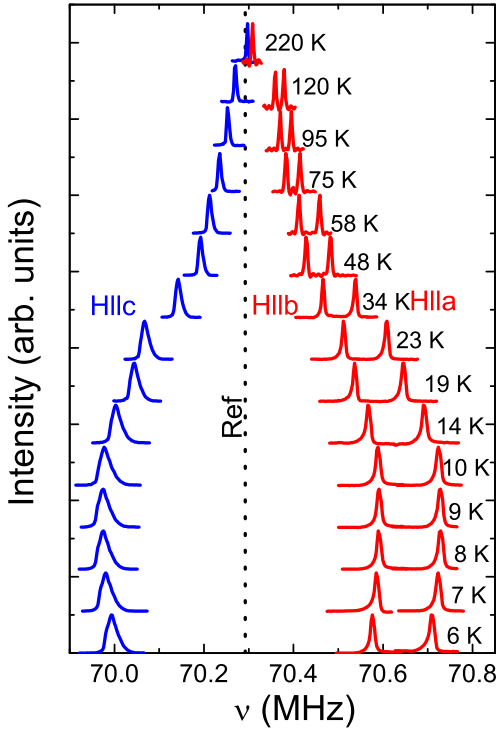}
\caption{\label{spk} (Color online) High-field Fourier
transform $^{31}$P NMR spectra for
the P1 site at different temperatures $T$ ($T > T_{\rm N}$) and along
different orientations for single crystalline Pb$_{2}$(VO)(PO$_{4}$)$_{2}$.
The vertical dotted line corresponds to the $^{31}$P resonance frequency 
of the reference sample H$_{3}$PO$_{4}$.}
\end{figure}
\section{\textbf{Results}}
\subsection{\textbf{NMR spectra above $T_{\rm N}$}}
\begin{figure}
\includegraphics {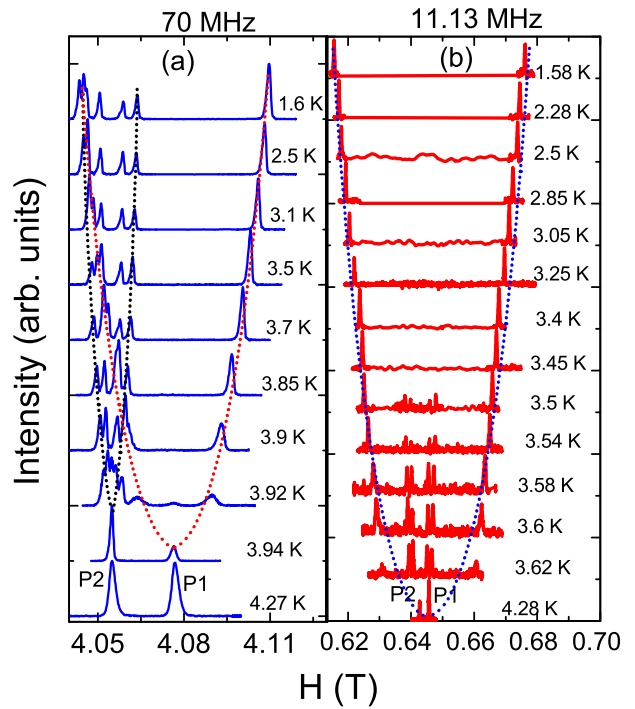}
\caption{\label{belowTn} (Color online) Temperature dependent $^{31}$P NMR
spectra measured at (a) 70.00~MHz and (b) 11.13~MHz for $H \parallel c$. 
The spectra in the paramagnetic state split below $T_{N}$ into two and 
four lines for the P1 and P2 sites, respectively, due to the internal 
field $H_{\rm int}$. The dashed lines are guides to the eye for the 
shifts of the satellite lines.}
\end{figure}

The $^{31}$P nucleus has nuclear spin $I=1/2$. For a $I=1/2$ nucleus 
one expects a single spectral line for each inequivalent P site.\cite{nath2005}
As shown in the crystal structure (Fig.~\ref{structure}),
Pb$_{2}$VO(PO$_{4}$)$_{2}$ has two inequivalent P sites.
We indeed observed two narrow spectral lines corresponding to
the two P sites for all three orientations of the crystal above 
$T_{\rm N}$ (e.g., see $^{31}$P NMR spectra at 4.27~K in Fig.~\ref{belowTn}).
For each P site the line position was found to shift with
temperature. However for the strongly coupled P1 site the 
shift was found to be much stronger compared to the weakly 
coupled P2 site and is also strongly orientation dependent. In the present 
work we have mainly focused on the P1 site and therefore data for the 
P2 site are not shown. Figure~\ref{spk} shows the $^{31}$P NMR
spectra for the P1 site measured on a single crystal
at different temperatures and field applied along different orientations.

\begin{figure}
\includegraphics [width=3.2in] {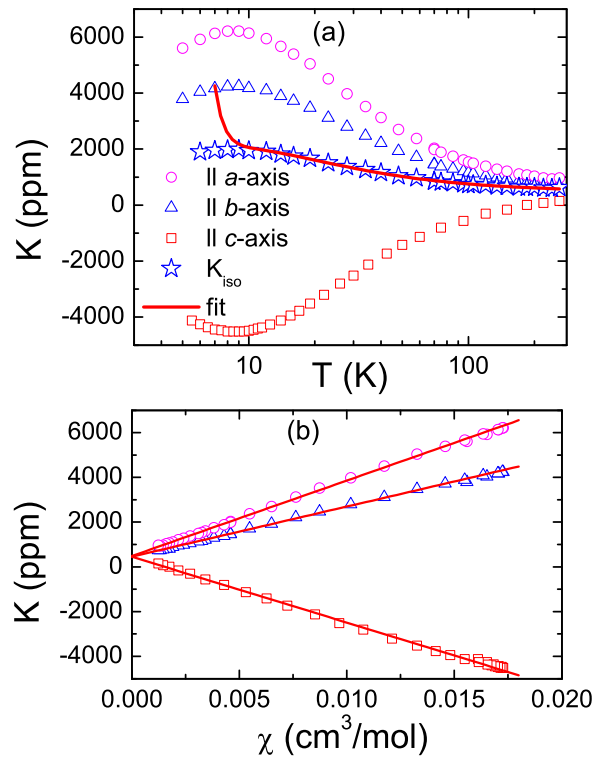}
\caption{\label{K} (Color online) (a) Temperature dependent NMR
shift $K$ vs. $T$ for the P1 site along different field orientations.  The
solid line is the fit of $K_{\rm iso}$ by Eq.~(\ref{shift}). (b)
$^{31}$P shift $K$ vs. $\chi$ measured at 4~T is plotted with
temperature as an implicit parameter for all three orientations. The
solid lines are linear fits.}
\end{figure}

Figure~\ref{K}(a) presents the temperature dependence of $K$ for the
P1 site derived from the data in Fig.~\ref{spk}. It shows a strong 
anisotropy along different directions.
Along the $a$- and $b$-axes it shows a strong positive shift while
along the $c$-axis it shows a strong negative shift. On the other
hand for the P2 site the shift is very weak.
The stronger and weaker shifts for the P1 and P2 sites,
respectively, suggest that the former one is strongly coupled while
the latter one is weakly coupled to the V$^{4+}$ spins,
consistent with expectation from the crystal structure. 
The temperature dependence
of the isotropic $^{31}$P1 NMR shift $K_{\rm iso}$ was calculated as
$K_{\rm iso}= (K_{a}+K_{b}+K_{c})/3$ and is also shown in
Fig.~\ref{K}(a). It shows a marked $T$ dependence
like $\chi (T)$.\cite{kaul2005} At high temperatures $K_{\rm iso}$ varies in
a Curie-Weiss manner and then passes through a broad maximum at around
9~K reflecting low-dimensional short-range antiferromagnetic ordering. Since
$\chi(T)$ for the single crystal at $T>T_{\rm N}$ is isotropic along
different orientations,\cite{kaul2005} this anisotropic $K(T)$ is likely due to
asymmetry in the hyperfine coupling constant between the V$^{4+}$ spins 
and the P nuclear spins.

\begin{figure}
\includegraphics [width=3.2in] {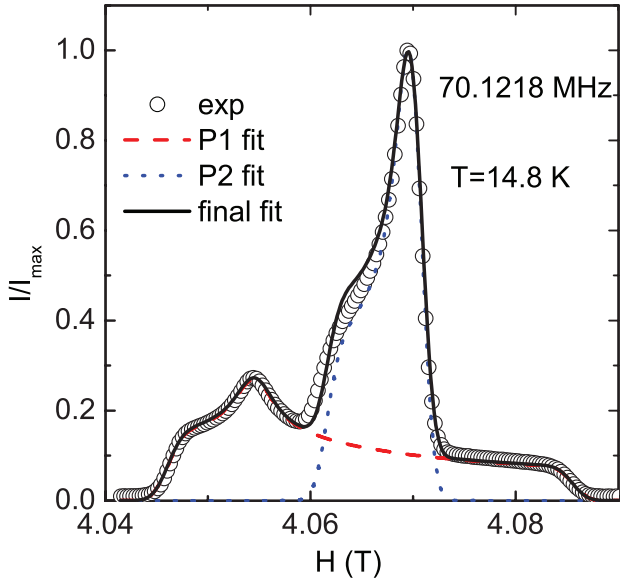}
\caption{\label{highfieldSPK} (Color online) The $^{31}$P NMR spectrum at
$14.8$~K measured at 70.1218~MHz on a polycrystalline Pb$_{2}$(VO)(PO$_{4}$)$_{2}$ 
sample (open circles). The dashed and dotted lines are the theoretical 
fits for the P1 and P2 sites, respectively, and the solid line is the 
total fit which is a superposition of the P1 and P2 fits. The NMR shift 
values along the $a$-, $b$-, and $c$-directions obtained from the fitting are 
($K_{a} \simeq 5500$~ppm, $K_{b} \simeq 3430$~ppm, and $K_{c} \simeq -4160$~ppm)
and ($K_{a} \simeq 1960$~ppm, $K_{b} \simeq -342$~ppm, and $K_{c} 
\simeq -512$~ppm) for the P1 and P2 sites, respectively.}
\end{figure}

A $^{31}$P spectrum on a polycrystalline sample is shown in 
Fig.~\ref{highfieldSPK}. Along with the most intense central 
line there is a broad background containing extra shoulder-like 
features on either side. In an attempt to fit the experimental
spectra we assumed that the central part with a small asymmetry
comes from the weakly coupled P2 site and the broader one with large
asymmetry is from the strongly coupled P1 site. In this way the superposition of
these two spectra gives a reliable fit as shown in Fig.~\ref{highfieldSPK}. 
One can see in Fig.~\ref{highfieldSPK} that for the broad spectrum 
there are three shoulder-like features clearly visible. These 
shoulder positions were found to shift with temperature. We took 
the derivative of the spectra and picked the point of inflection 
for each shoulder as a function of temperature (not shown) and 
they overlap nicely with the single crystal $K(T)$ confirming that 
the shoulders are due to anisotropic shift at the P1 site. On the 
other hand the narrow line (P2 site) was found to shift very weakly 
with temperature.

The NMR shift is a
direct measure of the spin susceptibility $\chi _{\rm spin}$, and 
quite generally $K (T)$ is written in terms of $\chi _{\rm spin}(T)$ 
as\cite{foot1}
\begin{equation}
K(T)=K_{0}+\frac{A_{\rm hf}}{N_{\rm A}} \chi_{\rm spin}(T),
\label{shift}
\end{equation}
where $K_{0}$ is the temperature-independent chemical (orbital) 
shift, $N_{\rm A}$ is the Avogadro number, $A_{\rm hf}$ is the 
hyperfine coupling constant between the P nuclear spins and the 
V$^{4+}$ electronic spins, and $\chi_{\rm spin}/N_{\rm A}$ is expressed 
in units of $\mu_{\rm B}$/Oe per electronic spin, which in our case 
means per formula unit. The conventional scheme for calculating 
$A_{\rm hf}$ is to take the slope of the $K$ vs $\chi _{\rm spin}$
plot with $T$ as an implicit parameter. Here we mainly focus on the
P1 site. For each orientation the $K$ vs.\ $\chi$ plot is fitted very 
well by a straight line [Fig.~\ref{K}(b)] over the whole temperature 
range ($T > T_{\rm N}$) yielding $A_{\rm hf} = (1882 \pm 40)$, 
($1251 \pm 42$), and $-(1642 \pm 55)$~Oe/$\mu_{\rm B}$ along
the $a$-, $b$-, and $c$-directions, respectively. The isotropic
hyperfine coupling constant is then calculated to be $A_{\rm iso}=
(A_{a}+A_{b}+A_{c})/3 = (497 \pm 46$)~Oe/$\mu_{\rm B}$.

In order to estimate the exchange couplings $J_{1}$ and $J_{2}$, we 
fitted the temperature dependence of $K_{\rm iso}$ above 15~K by 
Eq.~(\ref{shift}), where $\chi_{\rm spin}$ is the the high-temperature series
expansion prediction for the molar spin susceptibility of the 
frustrated square lattice model given by\cite{rosner2003}
\begin{equation}
\chi_{\rm spin}(T)=\frac{N_{A}g^{2}\mu_{B}^{2}}{k_{B}T}\sum_{n}
\left(\frac{J_{1}}{k_{B}T}\right)^{n}\sum_{m}c_{m,n}\left(
\frac{J_{2}}{J_{1}}\right)^{m},
\label{HTSE}
\end{equation}
where $c_{m,n}$ are the coefficients listed in Table~I of 
Ref.~\onlinecite{rosner2003}, $g$ is the $g$-factor, and 
$k_{\rm B}$ is the Boltzmann constant. From the electron paramagnetic 
resonance measurements at room temperature, $g$ parallel ($g_{\parallel}$) 
and perpendicular ($g_{\perp}$) to the $c$-axis were found to be 1.929 
and 1.966, respectively.\cite{foerster2009} The isotropic 
$g=\sqrt{(g^{2}_{\parallel}+2g^{2}_{\perp})/3}$ was calculated 
to be 1.95 which was kept fixed during the fitting procedure.
In this way we obtained $K_{0} = (462 \pm 11)$~ppm, $A_{\rm hf} 
= (481 \pm 20)$~Oe/$\mu_{\rm B}$, $J_{1} = (-5.4 \pm 0.5)$~K, 
and $J_{2} = (9.3 \pm 0.6)$~K\@. The fit is shown in 
Fig.~\ref{K}(a) as a solid line. These values of $J_{1}$ 
and $J_{2}$ are close to the previously reported values 
estimated from $\chi(T)$ analysis\cite{kaul2004,kaul2005}
and also consistent with the saturation field obtained from high-field 
magnetization isotherm measurements.\cite{tsirlin2009a} The estimated
$A_{\rm hf}$ matches with the $A_{\rm iso}$ calculated above from 
the $K$-$\chi$ analysis within the error bars.

\begin{figure}
\includegraphics {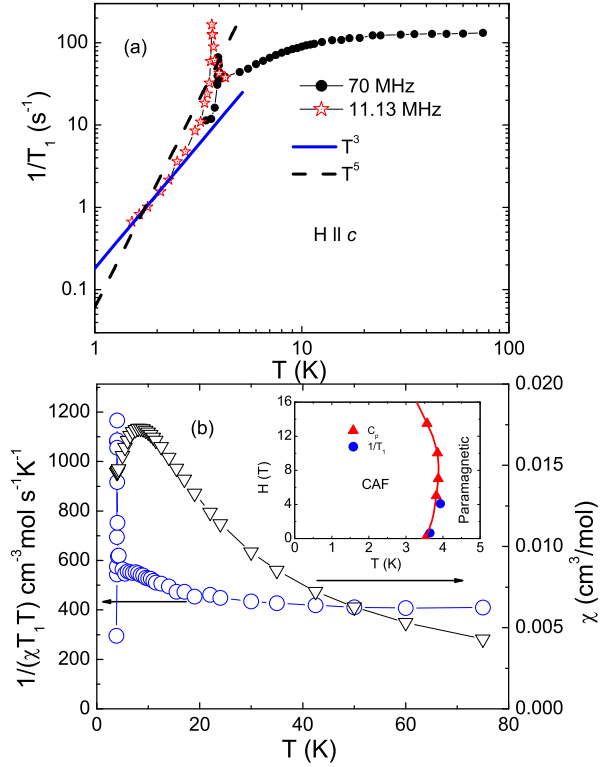}
\caption{\label{t1} (Color online) (a) Nuclear spin-lattice relaxation 
rate $1/T_{1}$ vs. temperature $T$ measured for the P1 site at 
70 and 11.13~MHz with the magnetic field applied along the $c$-axis. 
Below $T_{\rm N}$ measurements were done at the central field 
corresponding to the right-hand-side satellite [see Fig.~\ref{belowTn}(b)].
The solid and dashed lines represent $T^3$ and $T^5$ behaviors, 
respectively. (b) $1/(\chi T_{1}T)$ (left $y$-axis) and $\chi$ (right $y$-axis)
plotted as a function of $T$. The inset shows the $H$ vs. $T$ 
phase diagram where $T_{\rm N}$ values obtained from our $1/T_{1}$ 
measurements at the P1 site and the values from heat capacity 
$C_{\rm p}$ measurements (Ref.~\onlinecite{kaul2005}) are plotted.}
\end{figure}

\subsection{Nuclear spin-lattice relaxation rate $1/T_{1}$}

The $^{31}$P nuclear spin-lattice relaxation rate $1/T_{1}$ was 
measured with the magnetic field applied parallel to the $c$-axis. 
For a $I=1/2$ nucleus the recovery of the longitudinal magnetization 
is expected to follow a single-exponential
behavior. In Pb$_{2}$VO(PO$_{4}$)$_{2}$, the recovery of the nuclear
magnetization after a saturation pulse was indeed fitted well by
the exponential function
\begin{equation}
1-\frac{M(t)}{M_{0}}=Ae^{-t/T_{1}},
\label{exp}
\end{equation}
where $M(t)$ is the nuclear magnetization at a time $t$ after
the saturation pulse and $M_{0}$ is the equilibrium nuclear magnetization.
The temperature dependence of $1/T_{1}$ measured for the P1 site 
for $H\parallel c$ is presented in Fig.~\ref{t1}(a).
At high temperatures ($T \gtrsim 20$~K), $1/T_{1}$ is
temperature independent. In the high temperature limit $T\gg J$,
a temperature-independent $1/T_{1}$ behavior is
typical when paramagnetic moments fluctuate fast and at
random.\cite{moriya1956} With decrease in temperature, $1/T_{1}$ 
decreases slowly for $T<15$~K and then shows a peak around 3.93~K. This
decrease is very similar to that observed previously in the cases of
the antiferromagnetic square lattices VOMoO$_{4}$ from $^{95}$Mo
NMR\cite{carretta2002b} and [Cu(HCO$_{2}$)$_{2}$.4D$_{2}$O] from $^{1}$H
NMR where the decrease of $1/T_{1}$ above $T_{\rm N}$ is explained by 
cancellation of the antiferromagnetic spin fluctuations at the positions of the probed nuclei.\cite{carretta2000}  Below the peak $1/T_{1}$ again decreases smoothly towards zero.

For the polycrystalline sample, $1/T_{1}$ measurements were not 
possible due to broad spectra and overlapping of signals from the 
two P sites.

\subsection{NMR spectra below $T_{\rm N}$}
Below $T_{\rm N}$ both the P1 and P2 lines were found to split
(Fig.~\ref{belowTn}) indicating that both P sites are
experiencing the static internal field in the ordered state.
The P1 site line splits into two satellite lines while the P2 site 
line splits into four satellite lines
as a result of the hyperfine field between the P nuclei and the
ordered V$^{4+}$ moments. For the 70~MHz measurements [Fig.~\ref{belowTn}(a)],
the left-hand-side satellite of the P1 site overlaps with the four split
satellites of the P2 site making it difficult to distinguish the peaks. It is
possible to separately measure the spectra for the two different sites
by measuring at low frequency and using a faster repetition rate since
the relaxation rates associated with the two sites are different (for 
the P1 site $T_{1}$ is shorter than for the P2 site).
Therefore we remeasured the spectra for the P1 site at 11.13~MHz using
a faster repetition rate and the results are presented in Fig.~\ref{belowTn}(b).

\begin{figure}
\includegraphics [width=3.2in] {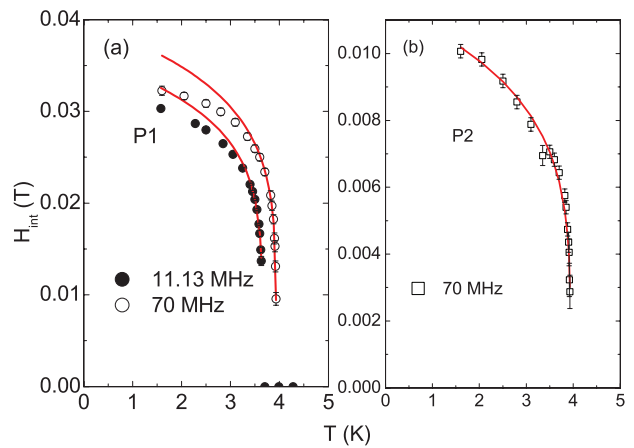}
\caption{\label{sublattice} (Color online) Temperature $T$ dependence 
of the internal field $H_{\rm int}$ obtained from NMR spectra for 
$H \parallel c$ in the ordered state (a) for the P1 site measured 
at 11.13 and 70 MHz and (b) for the P2 site measured at 70 MHz. 
$H_{\rm int}$ is proportional to the V$^{4+}$ sublattice 
magnetization. The solid lines are fits by Eq.~(\ref{ms}) to the data, 
as described in the text.}
\end{figure}

The internal field $H_{\rm int}$, which is proportional to the V$^{4+}$
sublattice magnetization, was determined from half the separation
between the split lines. The temperature dependences of $H_{\rm int}$ 
are shown in Fig.~\ref{sublattice} for both the P-sites measured for 
$H \parallel c$. For the P1 site at 70~MHz, the left-hand-side satellite 
was chosen as the peak of equal intensity to the right-hand-side 
satellite. The $H_{\rm int}(T)$ for the P1 site was found to be 
almost independent of field below $T_{\rm N}$. In order to extract the critical 
exponent of the order parameter (sublattice magnetization), $H_{\rm int}(T)$ 
was fitted by the power law
\begin{equation}
H_{\rm int}(T)=H_{0}\left(1-\frac{T}{T_{\rm N}}\right)^{\beta}.
\label{ms}
\end{equation}
One can notice that $H_{\rm int}$ decreases sharply on approaching $T_{\rm N}$.
For an accurate determination of the critical exponent $\beta$, one needs
data points close to $T_{N}$ (i.e.\ in the critical region). 
We have estimated $\beta$ by fitting the data points in temperature steps 
of 0.01~K as close as possible to $T_{\rm N}$ as shown in Fig.~\ref{sublattice}. 
For the P1 site at 11.13~MHz, the maximum value of $\beta = 0.25 \pm 0.02$ 
with $T_{\rm N} \simeq 3.655$~K was obtained by fitting the data points 
(3.58~K to 3.63~K ) close to $T_{\rm N}$. By increasing the number of 
fitting points towards low-$T$s, the $\beta$ value was found to decrease 
and the minimum value of $\beta = 0.15 \pm 0.02$ with $T_{\rm N} \simeq 3.64$~K 
was obtained by fitting the whole temperature range (1.58~K to 3.63~K). 
Similarly for the P1 site at 70~MHz, the maximum value of 
$\beta = 0.23 \pm 0.03$ with $T_{\rm N} \simeq 3.933$~K was obtained 
by fitting data points (3.9~K to 3.93~K) close to $T_{\rm N}$ 
and it decreases when low-$T$ data points were included. At low-$T$s, 
$H_{\rm int}$ develops the tendency of saturation and it saturates much 
faster than expected from mean field theory [see the deviation of fits 
in Fig.~\ref{sublattice}(a) at low-$T$s]. On the other hand for the 
P2 site at 70~MHz, close to $T_{\rm N}$, due to overlapping of 
the lines from the two P sites an accurate determination of $H_{\rm int}$ 
was not possible; therefore the data points shown in Fig.~\ref{sublattice}(b) 
for the P2 site have large error bars.

\begin{figure}
\includegraphics [width=3.2in] {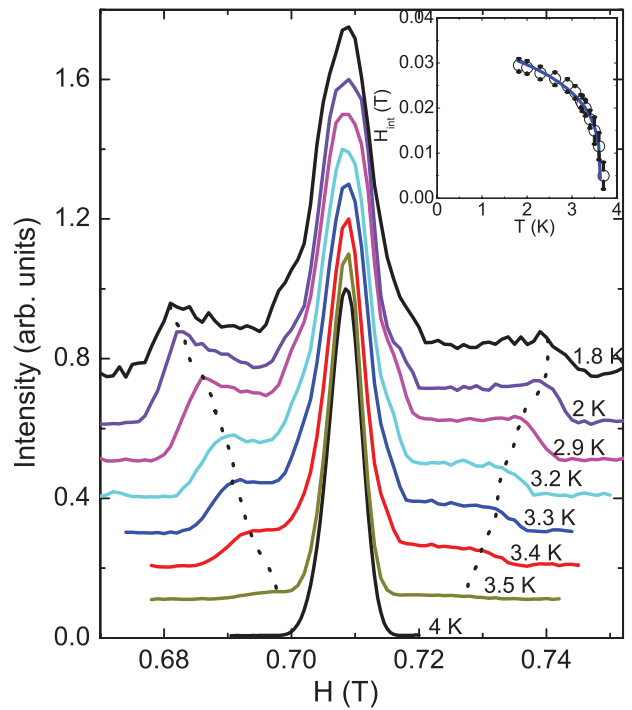}
\caption{\label{lowfieldSPK} (Color online) Temperature-dependent
$^{31}$P NMR spectra measured at $12$ MHz on a polycrystalline sample
below $T_{\rm N}$. The dotted lines indicate the splitting of the P1 line.
Inset: Internal field $H_{\rm int}$ as a function of temperature $T$. 
The solid line is a fit by Eq.~(\ref{ms}).}
\end{figure}

For the polycrystalline sample below $T_{N}$, the NMR line at 70~MHz 
was found to broaden abruptly. In order to check whether any extra 
features could be resolved, we remeasured the spectra at a low 
frequency (12~MHz) and again with a fast repetition rate. It was 
found that above $T_{N}$ the line remains narrow and immediately 
below $T_{N}$ it starts broadening (see Fig.~\ref{lowfieldSPK}). The 
broad line arising from the P1 site develops two satellites on 
either side of the central peak that are due to the P2 site. With a decrease in 
temperature these two satellites move away from each other.
\cite{vonlanthen2002} We then plotted half the distance between 
the two satellite positions ($H_{\rm int}$) as a function of 
temperature (inset of Fig.~\ref{lowfieldSPK}). Fitting the 
data from 1.7~K to 3.6~K by Eq.~(\ref{ms}) yielded $\beta = 0.25 \pm 0.07$.

\section{Discussion}
\subsection{Statics}
The exchange couplings extracted from our NMR iso-shift data 
are consistent with the values reported from $\chi(T)$ 
analysis.\cite{kaul2004,kaul2005} Based on the $J_{2}/J_{1}$ 
ratio, Pb$_{2}$VO(PO$_{4}$)$_{2}$ with $\alpha \simeq -1.72$ 
is located in the CAF regime of the phase diagram in 
Fig.~\ref{phasediagram}. In the crystal structure, squares
are formed via a V-O-P1-O-V superexchange interaction path. 
Since P1 is strongly coupled to the V$^{4+}$ spins via hyperfine 
interaction, any change in the structure properties would be 
reflected in the $^{31}$P NMR. The total hyperfine coupling constant 
at the P1 site is generally the sum of transferred hyperfine 
($A_{\rm trans}$) and dipolar ($A_{\rm dip}$) couplings produced 
by V$^{4+}$ spins, i.e.\ $A_{\rm hf}=z^{'}A_{\rm trans}+A_{\rm dip}$ 
where $z^{'}$ is the number of nearest-neighbor V$^{4+}$ spins of the P1 site. 
The anisotropic dipolar couplings were calculated for three orientations 
of the crystal using lattice sums with an assumption of 1~$\mu_{\rm B}$ magntic moment on each V$^{4+}$ ion, 
which are $A_{\rm dip}^{a}=150$~Oe/$\mu_{\rm B}$, $A_{\rm dip}^{b}
=170$~Oe/$\mu_{\rm B}$, and $A_{\rm dip}^{c}=-320$~Oe/$\mu_{\rm B}$ along
the $a$-, $b$-, and $c$-directions, respectively.  Thus by
subtracting the dipolar coupling from the total hyperfine coupling, 
the transferred hyperfine coupling was obtained to be $A_{\rm trans}=1730$, 
1080, and $-1320$~Oe/$\mu_{\rm B}$, respectively, suggesting that the 
dominant contribution to the total hyperfine coupling is due to the 
transferred hyperfine coupling at the P1 site.  The transferred hyperfine 
coupling at the P1 site arises mainly from interactions with the four 
nearest neighbor V spins in the plane. The isotropic and anisotropic 
transferred hyperfine couplings originate from P(3$s$)-O(2$p$)-V(3$d$) 
and P(3$p$)-O(2$p$)-V(3$d$) covalent bonds, respectively. Since P1 
is surrounded by four V ions forming an approximate square
 lattice in the plane, the experimentally observed asymmetry in total
 hyperfine field in the plane indicates inequivalent P(3$p$)-O(2$p$)-V(3$d$)
 bonds for the four nearest-neighbor V ions and hence a distortion
in the square lattice, consistent with the low symmetry of the crystal 
structure as pointed out in Sec.~I. Our observations seem to be
consistent with the recent theoretical calculation by Tsirlin et
al.\cite{tsirlin2009} where they have estimated the AFM NNN exchange 
couplings $J_{2}^{'}$ and $J_{2}^{''}$. The ratio is $J_{2}^{''}/J_{2}^{'} =0.67$, smaller than the value of unity expected for a regular frustrated square lattice, consistent with a distortion in the square.  Theoretical studies have indicated that the frustrating next-nearest-neighbor coupling $J_2$ can lead to a lattice distortion due to spin-lattice coupling.\cite{Becca2002, Weber2005}

Now we discuss the spin structure in the magnetically ordered state.
As shown in Fig.~\ref{belowTn}, for $H \parallel c$, the P1 line splits 
into two lines rather than being simply shifted to lower field in 
the antiferromagnetically ordered state. We also carried out spectral 
measurements below $T_{\rm N}$ for the field applied along $a$- 
and $b$-axes at 11.6~MHz. For both directions, the NMR line 
was also found to split but the splitting is much weaker compared to that with the field along the 
$c$-axis. For instance, at $T \approx 2.9$~K, the internal field 
$H_{\rm int}$ along the $a$- and $b$-axes are $\sim 5$~Oe and 
$\sim 10$~Oe, respectively, while along the $c$-axis it is about 
250~Oe.\cite{foot4} Thus the direction of 
the internal field $H_{\rm int}$ at the P1 site is almost parallel 
(or antiparallel) to the applied field $H$ along the $c$-axis. 
In this case the effective
field at the P1 site would be $H_{\rm eff}=H \pm H_{\rm int}$. The satellite 
lines are rather sharp with no striking broadening. Thus the splitting of 
the NMR line with no broadening is direct evidence of a commensurate 
magnetic structure.  If the magnetic structure were incommensurate with 
the lattice, the internal field would be distributed and the spectrum 
would not exhibit sharp resonance lines as seen in Fig.~\ref{belowTn}. 
As pointed out before by neutron scattering experiments,
a diffuse magnetic background reminiscent of quantum disorder was
observed while approaching $T_{\rm N}$.\cite{skoulatos2007} In NMR this 
magnetic background would have appeared as line broadening. Thus our 
experimentally observed narrow satellites with no further splitting 
do not indicate the presence of magnetic/structural disorder, which 
is inconsistent with the interpretations of $\mu$SR 
(Ref.~\onlinecite{carretta2009}) and neutron scattering 
(Ref.~\onlinecite{skoulatos2007}) results measured on powder 
samples. At present the origin of this discrepancy is not clear to us. 
Further inelastic neutron scattering and $\mu$SR experiments on 
single crystals are required for a better understanding of this issue.

Our observations are very similar to $^{75}$As NMR measurements on 
(Ba,Sr,Ca)Fe$_{2}$As$_{2}$ in the stripe/columnar type magnetically 
ordered state.\cite{kitagawa2008,kitagawa2009,baek2009}
According to Kitagawa et al.,\cite{kitagawa2008} based on the crystal 
symmetry of (Ba,Sr,Ca)Fe$_{2}$As$_{2}$ with orthorhombic structure 
($Fmmm$), the internal field at the $^{75}$As-site is
explained by an off-diagonal transferred hyperfine field induced by four 
nearest-neighbor Fe moments whose direction is in the
$ab$-plane. In the case of N\'{e}el type AFM order, the $c$-component 
of $H_{\rm int}$ at the As site is zero due to a perfect 
cancellation of the off-diagonal hyperfine fields produced by 
the four Fe moments when the spin moments are in the $ab$-plane. 
Only the stripe-type AFM order can produce a $c$-component of
$H_{\rm int}$ at the As site.
In the case of Pb$_2$VO(PO$_4$)$_2$, the crystal structure is monoclinic
($P2_{1}/a$) with a lower symmetry. Thus the relationship between $H_{\rm int}$
and the magnetic structure is not trivial. In addition, the hyperfine
field at the P1 site contains both transferred and dipolar
components. However, it may be possible to consider the present
case based on the discussion for (Ba,Sr,Ca)Fe$_{2}$As$_{2}$ systems as a
first approximation because the local symmetry in the plane is 
close to four-fold axial symmetry corresponding
to a square lattice formed by V ions with respect to the P1 site. Since the
magnetic easy axis is the $b$-axis and the direction of $H_{\rm int}$ is along
the $c$-axis in the magnetic ordered state, it is concluded that magnetic order
in Pb$_2$VO(PO$_4$)$_2$ has a columnar structure with spins in the
$ab$-plane. Our interpretation in terms of a columnar magnetic structure is consistent with the neutron scattering experiments.\cite{skoulatos2009} Of course, as we pointed out, the four-fold symmetry of the square lattice is slightly broken due to the lower symmetry of the crystal structure which makes the analysis more complicated. The small $a$- and $b$-components of 
$H_{\rm int}$ might originate from the non-perfect cancellation 
of the hyperfine field due to the low symmetry crystal structure. 
The observed splitting to four lines of the P2 site 
spectrum could also be related to such a complication.

\subsection{Dynamics}

The overall temperature dependent behaviors of $1/T_{1}$ for both P
sites are alike. As shown in Fig.~\ref{t1}(b), $1/(\chi T_{1}T)$ above
$\sim 20$~K is $T$-independent and increases slowly below 20~K where the
system begins to show strong antiferromagnetic short-range correlations. In the same figure $\chi(T)$ is also plotted to highlight the broad maximum
corresponding to the short-range correlations.  In general, $1/(T_{1}T)$
is expressed in terms of the dynamic susceptibility $\chi_{M}(\vec{q},
\omega_{0})$ per mole of electronic spins as\cite{moriya1963,foot3,mahajan1998}
\begin{equation}
\frac{1}{T_{1}T} = \frac{2\gamma_{N}^{2}k_{B}}{N_{\rm A}^{2}}
\sum\limits_{\vec{q}}\mid A(\vec{q})\mid
^{2}\frac{\chi^{''}_{M}(\vec{q},\omega_{0})}{\omega_{0}},
\label{t1form}
\end{equation}
where the sum is over wave vectors $\vec{q}$ within the first Brillouin zone, 
$A(\vec{q})$ is the form factor of the hyperfine interactions as a
function of $\vec{q}$ in units of Oe/$\mu_{\rm B}$, and 
$\chi^{''}_{M}(\vec{q},\omega _{0})$ is the imaginary part of the 
dynamic susceptibility at the nuclear Larmor frequency $\omega _{0}$ 
in units of $\mu_{\rm B}$/Oe. The uniform static molar
susceptibility $\chi=\chi_{M}^{'}(0,0)$ corresponds to the real component 
$\chi_{M}^{'}(\vec{q},\omega _{0})$ with $q=0$ and $\omega_{0}=0$. Thus
the temperature-independent behavior of $1/(\chi T_{1}T)$ above 20~K 
in Fig.~\ref{t1}(b) demonstrates that the $T$-dependence of
$\sum\limits_{\vec{q}}\mid A(\vec{q})\mid
^{2}\chi^{''}_{M}(\vec{q},\omega _{0})$ scales to that of $\chi$.
On the other hand, a slight increase of $1/(\chi T_{1}T)$
below 20~K indicates that $\sum\limits_{\vec{q}}\mid A(\vec{q})\mid
^{2}\chi^{''}_{M}(\vec{q},\omega _{0})$ increases more
than $\chi$ due to the growth of antiferromagnetic correlations
with decreasing $T$. There could be two possible sources for this 
growth of antiferromagnetic correlations. One is N\'{e}el type antiferromagnetic 
spin fluctuations characterized by the wave vector $\vec{q}= 
(\pm \pi/a, \pm \pi/b)$ and the other is columnar type AFM spin fluctuations with $\vec{q}=(\pm \pi, 0)$ above $T_{\rm N}$. The hyperfine form factor at the P1 site can be written as ${|A(\vec{q})|}^{2}=\{2A[\cos(q_{x}a/2) + \cos(q_{y}b/2)]\}^2$,
if P1 is located at the center of the perfect square lattice
formed by V ions. In such a case, N\'{e}el type antiferromagnetic 
spin fluctuations do not contribute to $1/(\chi T_{1}T)$ since 
${|A(\vec{q})|}^{2}$ is zero. However, in Pb$_{2}$VO(PO$_{4}$)$_{2}$, 
since P1 is not located exactly at the center of a square lattice, 
$A(\vec{q})$ should have a nonzero value. Therefore the enhancement of
$1/(\chi T_{1}T)$ could be due to the growth of N\'{e}el type AF spin 
fluctuations which are not completely filtered out at the P1 site. 
Alternatively, since the AFM ordered state is columnar type as discussed 
before, if it has a finite correlation extended above $T_{\rm N}$ 
then ${|A(\vec{q})|}^{2}$ would have a finite value which might 
also be responsible for the enhancement of $1/(\chi T_{1}T)$.  If large enough single crystals could be grown, the origin of the enhancement could be conclusively resolved from inelastic neutron scattering measurements of the magnetic fluctuation wavevector above $T_{\rm N}$.

A sharp peak in $1/T_{1}$ versus $T$ is a direct indication of long-range magnetic ordering.  However, the peak position for the $H = 0.65$~T measurement is at 3.65~K while for 4.06~T it is enhanced to 3.93~K\@. We plotted these two points in the $H$~vs.~$T_{\rm N}$ phase diagram reported in Ref.~\onlinecite{kaul2005}
obtained from heat capacity measurements of $T_{\rm N}(H)$ [see inset of
Fig.~\ref{t1}(b)], where it is clearly seen that $T_{\rm N}$ increases
slightly with increasing $H$ from 0 to 7~T
and then decreases with further increase of the field.  This behavior is 
similar to that observed in FSL compound BaCdVO(PO$_{4}$)$_{2}$ 
(Ref.~\onlinecite{nath2008}) 
as well as in non-frustrated 2D antiferromagnet Mn(HCOO)$_{2} \cdot$2H$_{2}$O 
[see p.~388 of Ref.~\onlinecite{jongh1989}].  Qualitatively,
 long-range magnetic ordering in low-dimensional and/or
frustrated spin systems is suppressed by quantum fluctuations.
Magnetic field suppresses these quantum fluctuations, therefore $T_{\rm N}$ is
slightly enhanced with increasing field. However, above 10~T the field is
strong enough to partially suppress the antiferromagnetic ordering, hence
$T_{\rm N}$ is reduced. In the antiferromagnetic ordered state, $1/T_{1}$ 
is mainly driven by scattering of magnons off nuclear spins, leading to 
a power law temperature dependence.\cite{beeman1968,belesi2006} For $T \gg \Delta$, 
where $\Delta$ is the gap in the spin-wave spectrum in temperature 
units, $1/T_{1}$ either follows a $T^{3}$ behavior due to a two-magnon 
Raman process or a $T^{5}$ behavior due to a three-magnon process, 
while for $T \ll \Delta$, it follows an activated behavior 
$1/T_{1} \propto T^{2}\exp(-\Delta/T)$. As seen from Fig.~\ref{t1}(a), 
 our $^{31}$P $1/T_{1}$ data in the lowest temperature region 
 (1.5~K $\leq T \leq$ 2.3~K) 
 follow a $T^{3}$ behavior rather than a $T^{5}$ behavior suggesting that the 
 relaxation is mainly governed by the two-magnon Raman process. The lack of 
 activated behavior down to 1.5~K indicates that the upper limit of 
 $\Delta$ is 1.5~K. Therefore, the upper limit of the axial anisotropy 
 [$D \approx \Delta^2/(|J_{1}|+|J_{2}|)$] (Ref.~\onlinecite{melzi2001}) would be about 0.15~K\@.

The $^{31}$P $1/T_{1}$ data and the hyperfine coupling constants 
determined above enable us to estimate the exchange
couplings in the system under investigation. According to Moriya, the nuclear
spin-lattice relaxation rate $1/T_{1}$ at sufficiently high temperature in a
system with exchange-coupled local moments is constant
and can be expressed within the Gaussian approximation of the
correlation function of the electronic spin as\cite{moriya1956, foot2}
\begin{eqnarray}
\left(\frac{1}{T_1}\right)_{T\rightarrow\infty} &=&
\frac{(\gamma_{N} g\mu_{\rm B})^{2}\sqrt{2\pi}z^{'}S(S+1)}{3\omega_{E}} \nonumber\\
&\times&
\Bigg[\frac{\Big(\frac{A_{hf}^{a}}{z^{'}}\Big)^{2}+\Big(\frac{A_{hf}^{b}}{z^{'}}\Big)^{2}}{2}\Bigg],
\label{t1inf}
\end{eqnarray}
where $\omega _{E}=[\max(|J_{1}|,|J_{2}|)k_{\rm B}/\hbar]\sqrt{2zS(S+1)/3}$\ 
is the Heisenberg exchange frequency, $z$
is the number of nearest-neighbor spins of each V$^{4+}$ ion, and $z^{'}$ 
is the number of nearest-neighbor V$^{4+}$ spins of the P1 site. 
In order to get the coupling of the P1 site to individual V$^{4+}$ spins, 
the total coupling is divided by $z^{'}$. Since our measurements were 
done for $H \parallel c$, the hyperfine couplings in the $ab$-plane 
(i.e.\ $A_{hf}^{a}$ and $A_{hf}^{b}$) contribute to $1/T_{1}$. 
Therefore, we have taken the rms average of couplings along the 
$a$- and $b$-directions. The $z^{'}$ in the numerator is due to 
the fact that the P1 site feels the fluctuations arising from 
all nearest-neighbor V$^{4+}$ spins. Using the relevant parameters 
($A_{hf}^{a} \simeq 1882$~Oe/$\mu_{\rm B}$, $A_{hf}^{b} \simeq 
1251$~Oe/$\mu_{\rm B}$, $z=4$, $z^{'}=4$, $S=1/2$, and the high 
temperature (25~K to 100~K) relaxation rate of 
$\left(\frac{1}{T_1}\right)_{T\rightarrow\infty}\simeq 127$~s$^{-1}$ for 
the P1 site) in Eq.~(\ref{t1inf}), the magnitude of the maximum 
exchange coupling constant was calculated to be 
$\max(|J_{1}|,|J_{2}|) \simeq 8$~K. For a system with two 
different exchange mechanisms, the exchange frequency is mainly dominated 
by the larger exchange interaction. Therefore, in Pb$_{2}$VO(PO$_{4}$)$_{2}$, 
this value of $J$ should represent the dominant exchange constant $J_{2}$. 
Indeed, $\max(|J_{1}|,|J_{2}|) \simeq 8$~K is close to the value 
$J_{2} = (9.3 \pm 0.6)$~K obtained from the $K(T)$ analysis above.

\subsection{Critical effects}
The NMR spectrum measurements can give a precise estimate of the 
temperature dependence of sublattice magnetization, which cannot be
obtained from static magnetization measurements. The temperature 
dependence of $H_{\rm int}$ can potentially provide the critical 
exponent $\beta$ reflecting the universality class/spin dimensionality 
of the spin system. In Pb$_{2}$VO(PO$_{4}$)$_{2}$, 
due to partial filtering of AF fluctuations above $T_{\rm N}$, our 
$1/T_{1}$ data restricted us from doing a detailed analysis. The $\beta$ 
values expected for different spin- and lattice-dimensionalities are 
listed in Table~\ref{beta}.\cite{collins1989,jongh1989,pelissetto2002,
ozeki2007,bramwell1993} If we assume that the experimental $\beta$ 
value of $\approx 0.23$ obtained above from the measurements on
the single crystal is due to critical fluctuations, then 
\begin{table}
\caption{\label{beta} Critical exponent $\beta$ expected for 
different universality classes.  The value of $\beta$ listed for the 2D~XY model is for a finite-size system.}
\begin{ruledtabular}
\begin{tabular}{ccc}
  & $\beta$ & Ref. \\ \hline
 3D Heisenberg & 0.367, 0.33, 35 & \onlinecite{collins1989, pelissetto2002, jongh1989} \\ \hline
 3D XY & 0.345, 0.31, 0.33 & \onlinecite{collins1989, ozeki2007, jongh1989} \\ \hline
 3D Ising & 0.326, 0.325, 0.31 & \onlinecite{collins1989, ozeki2007, jongh1989} \\ \hline
 2D XY & 0.231 & \onlinecite{bramwell1993} \\ \hline
 2D Ising & 1/8 & \onlinecite{ozeki2007, collins1989} \\
\end{tabular}
\end{ruledtabular}
\end{table}
the $\beta$ value is closer 
to the 2D XY than the 3D models although the ordering is suggested to be a 3D 
order from the powder neutron diffraction measurements.\cite{skoulatos2009} 
We suggest that our data are not sufficiently close to $T_{\rm N}$ to reflect 
3D critical behavior.\cite{jongh1989} It is to be noted that the longitudinal 
$\mu$SR relaxation rate above $T_{\rm N}$ in Pb$_{2}$VO(PO$_{4}$)$_{2}$ has 
been fitted well by this 2D XY model.\cite{carretta2009} A tiny in-plane 
anisotropy can dominate the fluctuations if the dominant AF correlation 
length is large.\cite{suh1995} In the present case, $\chi(T)$ above 
$T_{\rm N}$ is approximately isotropic, but it is possible that there 
might be a small anisotropy which is not visible in the experimental 
$\chi(T)$. In fact, strong in-plane correlation may also enhance 3D coupling
along the $c$-direction.\cite{ding1992} Therefore to understand this 
issue, a precise estimation of in-plane spin anisotropy is required.

It is useful to compare our data with data for other layered
compounds. A similar reduced value of $\beta$ close to the $\beta$ for the 2D Ising model has been observed before in several layered compounds, e.g., 
K$_{2}$NiF$_{4}$, K$_{2}$MnF$_{4}$, K$_{2}$CoF$_{4}$, Rb$_{2}$MnF$_{4}$, 
and RbCoF$_{4}$ (Ref.~\onlinecite{jongh1989}). In these compounds 
the weak Ising-type anisotropy is about two orders of magnitude 
larger than the interplanar coupling.  Therefore it 
is argued that the long-range ordering is primarily induced by the strong 
anisotropy crossover from 2D Heisenberg to 2D Ising behavior, which precedes 
the 2D Ising to 3D Ising crossover that should occur close to $T_{\rm N}$. 
Since the onset of such 3D ordering is still governed by the 2D processes, 
the phase boundaries that are observed should reflect the underlying 2D 
character. Whenever there is a 3D ordering, one must have 3D correlations 
sufficiently close to $T_{\rm N}$. For instance, in BaNi$_{2}$(PO$_{4}$)$_{2}$, 
an exponent ($\beta=0.33$) close to 3D model was found near $T_{\rm N}$ and 
another exponent ($\beta=0.23$) close to that of the 2D XY model was found a little bit 
away from $T_{\rm N}$ (see p.~281 of Ref.~\onlinecite{jongh1989}). However, 
in some cases this 3D critical region is over such a narrow temperature 
range that it is not accessible experimentally. Such a scenario has been 
realized before in the frustrated antiferromagnetic square lattice 
compound Li$_{2}$VOSiO$_{4}$. In this compound, neutron powder-diffraction 
experiments show that the ordering is 3D and 
the antiferromagnetic $ab$-layers are coupled ferromagnetically 
along the $c$-axis.\cite{bombardi2004} In contrast, a reduced
value of $\beta$ close to the $\beta$ value of the 2D XY model has been reported from NMR measurements and the authors suggested that 
the transition to the columnar phase might be driven by the XY 
anisotropy.\cite{melzi2001,melzi2000} Although our experimental
observations below $T_{\rm N}$ are similar to those reported for
Li$_{2}$VOSiO$_{4}$, a proper explanation
for this peculiar behavior is still lacking in all these vanadates.
Therefore further experiments on high quality single crystalline 
materials are essential to address this issue.

\section{\textbf{Conclusion}}

We performed $^{31}$P NMR measurements on Pb$_{2}$VO(PO$_{4}$)$_{2}$, 
which is a strongly frustrated 2D Heisenberg square lattice compound. 
The exchange couplings were estimated reliably from $K(T)$ analysis to be
$J_{1} =(-5.4\pm 0.5)$~K and $J_{2} =(9.3\pm 0.6)$~K\@.
From the NMR spectral measurements, the ground state was detected to be
columnar antiferromagnetic (CAF) type and is consistent with that expected 
from the phase diagram (Fig.~\ref{phasediagram}) for 
$\alpha \equiv J_{2}/J_{1} \simeq -1.72$. The hyperfine
coupling and $1/T_{1}$ above $T_{\rm N}$ are consistent with the 
distortion of the V$^{4+}$ squares due to monoclinic crystal symmetry. 
We have measured the temperature dependence of the sublattice 
magnetization in the CAF ordered state from the line splitting. 
The critical exponent $\beta$ estimated from the sublattice 
magnetization was close to the value predicted for a 2D XY model. 
We did not observe any signature of magnetic/structural disorder in 
contrast to the neutron scattering and $\mu$SR experiments on
powder samples.\cite{carretta2009,skoulatos2007,skoulatos2009} 
To gain more insight, further complementary experiments on 
single crystals are required. Due to the low energy scale of 
the exchange interactions, this compound is also a potential candidate
for high-field experiments, and for high-pressure experiments where 
one can tune the exchange couplings.

\begin{acknowledgments}
We thank P. Carretta, A. A. Tsirlin and F. Becca for fruitful discussions and correspondence.  Work at the Ames 
Laboratory was supported by the Department of
Energy-Basic Energy Sciences under Contact No.~DE-AC02-07CH11358.
\end{acknowledgments}


\begin{thebibliography}{99}
\bibitem{foot1} Most people incorrectly write $K$ in terms of 
$\chi_{\rm spin}$ (cm$^{3}$/mol) 
as $K(T)=K_{0}+\frac{A_{\rm hf}}{N_{A}\mu_{\rm B}} \chi_{\rm spin}(T)$ 
with the units of $A_{\rm hf}$ expressed in Oe/$\mu_{\rm B}$. In order 
to match units between the left- and right-hand sides of this expression, the 
units of $A_{\rm hf}$ should not be Oe/$\mu_{\rm B}$, but rather Oe only. 
However, since Oe is the unit of hyperfine field and not of the 
hyperfine coupling constant, in order to have the correct units of 
$A_{\rm hf}$(Oe/$\mu_{\rm B}$), the above expression 
should instead be written as $K(T)=K_{0}+\frac{A_{\rm hf}}{N_{A}} 
\chi_{\rm spin}(T)$, with $\chi_{\rm spin}$ in units of $\mu_{\rm B}$/(Oe mol).
\end{thebibliography}
\end{document}